\documentclass[runningheads,fleqn]{svmult}

\usepackage{makeidx}   
\usepackage{graphicx}  
\usepackage{subeqnar}  
\usepackage{multicol}  
\usepackage{physproc}  
\makeindex             

\begin{document}
\title*{Hamiltonian structure of 2+1 dimensional gravity\protect\newline}
\toctitle{Focusing of a Parallel Beam to Form a Point
\protect\newline in the Particle Deflection Plane}
%
%
\titlerunning{2+1 dimensional gravity}
%
\author{Pietro Menotti
}
%
%
\institute{Dipartimento di Fisica della Universit\`a, 56100 Pisa,
Italy and \\
INFN Sezione di Pisa, 56100 Pisa, Italy}

\maketitle              

\begin{abstract}
A summary is given of some results and perspectives of the hamiltonian
ADM  
approach to $2+1$ dimensional gravity. After recalling the classical
results for closed universes in absence of matter we go over the the
case in which matter is present in the form of point spinless
particles. Here the maximally slicing gauge proves most effective by
relating $2+1$ dimensional  gravity to the Riemann- Hilbert problem. 
It is possible to solve the gravitational field in terms of the
particle degrees of freedom thus reaching a reduced dynamics which
involves only the particle positions and momenta. Such a dynamics is
proven to be hamiltonian and the
hamiltonian is given by the boundary term in the gravitational action.  
As an illustration the two body hamiltonian is used to provide the
canonical quantization of the two particle system.

\bigskip\bigskip\bigskip

\end{abstract}

\section{Introduction}
%

The past fifteen years have witnessed a remarkable interest in
gravity in 2+1 dimensions \cite{DJH} both at the classical and quantum
level. 

In this paper we shall summarize some old results and some recent
developments with regard to the hamiltonian formulation of the theory. We
shall insist on the conceptual side; the technical details can be found
in the published papers and reports.

\section{2+1 dimensional gravity in absence of matter}

In absence of matter the only degrees of freedom are given by the
moduli of the space sections; thus the open  universe case and the
case of the sphere topology are trivial. 

In absence of boundaries the action of the gravitational field reduces
to the Einstein- Hilbert term which can be put in hamiltonian form as
\begin{equation}\label{hilbertaction}
S_H = \int dt \int_{\Sigma_t} d^Dx
\left[ \pi^{ij}\dot g_{ij} - N^i H_i - NH\right]
\end{equation}
where we used the standard ADM metric \cite{ADM} 
\begin{equation}\label{ADMmetric2}
ds^2= -N^2 dt^2+ 2 g_{z\bar z}(dz+ N^z dt)(d\bar z+ N^{\bar z} dt).
\end{equation}
The choice of the gauge is of crucial importance in dealing with the
problem. The well known York gauge in which the time slices are
provided by the $D$ (in our case $D=2$) dimensional surfaces with $K=
{\rm const.}$, being $K$ the trace of the intrinsic curvature tensor
is particularly powerful as this gauge decouples the solution of the
diffeomorphism  constraint from that of the hamiltonian
constraint. Exploiting this feature it is possible \cite{moncrief,hosoya} 
to solve the diffeomorphism constraint and to provide
the hamiltonian on the reduced phase space given by the
moduli and their conjugate momenta. The number of moduli are $6g-6$ for
genus $g$ larger than $1$ and $2$ for the torus topology. The explicit
computation of the 
hamiltonian can be performed only in the simplest case of torus
topology. It is given by
\begin{equation}
H = \sqrt{\tau^2_2(p_1^2+p_2^2)}.
\end{equation}    
Quantization proceeds \cite{hosoya2,carlip} by replacing the canonical
variables with operators 
according to the correspondence principle. The ordering problem always
subsists; the most natural ordering translates the classical
hamiltonian into the square root of the Maass laplacian \cite{maass}
giving rise to 
the Schroedinger equation
\begin{equation}
i\frac{\partial\psi(x,y,t) }{\partial t}= \sqrt{-y^2(\partial_x^2+
\partial_y^2)}~\psi(x,y,t). 
\end{equation}    
The Maass laplacian has been widely investigated by mathematicians
\cite{fay,terras,puzio}. The classical as well the quantum hamiltonians
are  
invariant under modular transformations which in the case of the torus
is given by the group $SL(2,Z)$ and thus the solutions should be
invariant under such modular transformations. The eigenvalue problem
under this condition is not trivial and the spectrum well studied
\cite{fay,terras,puzio}. Such an approach gives a complete quantum
treatment of universes without matter with torus topology in the York
gauge.

\section{2+1 dimensional gravity coupled to particles}

We come now to the more difficult and realistic case of gravity
coupled to particles. It describes also a situation
of 3+1 dimensional gravity, i.e. the interaction of parallel cosmic
strings. 

One starts from the action of the gravitational field coupled to a
finite number of spinless point particles. The inclusion of spin to
point particles is not a trivial issue as it gives rise to closed time
like curves. 

We must add to the action (\ref{hilbertaction}) the particle action 
\begin{equation} 
S_m=\int\!d t \sum_n\Big(P_{ni}\, \dot
q_n^i+N^i(q_n) P_{ni} - N(q_n) \sqrt{P_{ni} P_{nj} g^{ij}(q_n)+
m_n^2}\Big)
\end{equation}
and for open universes we must also add boundary terms; we shall
adopt here the so called 
Trace-$K$ form completed of the terms which lye on two sub-manifolds of 
dimension $D-1$, being $D+1$ the dimension of space- time
\cite{hawkinghunter,brownyork}. In hamiltonian form it reduces to
\begin{equation}\label{boundaryaction} 
S_B = - \int dt H_B 
\end{equation} 
with
$$
-H_B= 2 \int_{B_t} d^{(D-1)}x \,\sqrt{\sigma_{Bt}} N 
\left( K_{B_t}+\frac{\eta}{\cosh\eta}{\cal D}_\alpha v^\alpha\right)
$$
\begin{equation}\label{boundaryhamiltonian} 
~~~~~~~~~~~~~~~-2 \int_{B_t} d^{(D-1)}x\, r_\alpha
\pi^{\alpha\beta}_{(\sigma_{Bt})} N_\beta. 
\end{equation}
The really important term turns out to be the first i.e. 
$\sqrt{\sigma_{Bt}} N K_{B_t}$ where  $K_{B_t}$ is the
extrinsic curvature of the $D-1$ dimensional boundary  (in our case
one- dimensional) of the time
slices as a sub-manifold embedded in the
$D$ dimensional time slices and $d^{(D-1)}x \sqrt{\sigma_{Bt}}$ is the
volume form induced by the space metric on the $D-1$ dimensional
boundary.  

We recall that the action $S_H+ S_B + S_m$ is so constructed as to provide
the correct 
equations of motion (i.e. Einstein's equations) when one computes a
stationary point of the action by keeping the values of the metric
fixed on the boundary. Such a procedure is equivalent \cite{wald} to the
weaker requirement of keeping fixed the intrinsic metric of the boundary.

One could in principle adopt also here the York
slicing. However the equations for the diffeomorphism and hamiltonian
constraints are not at all trivial. In particular the hamiltonian
constraint gives rise to an equation more complex than the
inhomogeneous sine-Gordon equation. Progress has been
achieved by the introduction of the instantaneous York gauge
\cite{BCV,welling,MS1,MS2,CMS}, or maximally slicing gauge. This is defined by
all time slices having 
$K=0$. Simple application of the Gauss- Bonnet theorem shows that such
a gauge can be applied only to open universes, or universes with the
topology of the sphere \cite{welling,MS1}.  In addition a closer
inspection shows that  
for the sphere topology it can describe only the simple stationary
case \cite{MS2}. Thus application of the $K=0$ gauge is practically
restricted 
to open universes, but here it proves very powerful. The technical
reason is the immediate solution of the diffeomorphism constraint 
given by
\begin{equation}\label{pibarzz}
\pi^{\bar z}_{~z} = -\frac{1}{2\pi}\sum_n\frac{P_n}{z-z_n}
\equiv -2\frac{\prod_B(z-z_B)}{{\cal P}(z)},
\end{equation}
where $z_n$ and $P_n$ are the complex positions and canonical
momenta of the particles, 
and
the reduction of the hamiltonian constraint to an inhomogeneous
Liouville equation, given by
\begin{equation}
\label{liouvilleeq}
2\Delta\tilde\sigma=-e^{-2\tilde\sigma}-4\pi \sum_n \delta^2(z- z_n)(
\mu _n -1)-4\pi\sum_B \delta^2(z- z_B),    
\end{equation}
to which powerful mathematical methods apply \cite{kra}. 
Here $\tilde\sigma$ is defined by
\begin{equation}\label{sigma}
e^{2\sigma} = 2 \pi^{\bar z}_{~z} \pi^{z}_{~\bar z} e^{2\tilde\sigma},
\end{equation} 
and is related to the space metric by $g_{ij}= \delta_{ij}e^{2\sigma}$.
In the above equation the sources are given by the particle and in
addition by the zeros of eq.(\ref{pibarzz}) here denoted by
$z_B$. These are 
the points where the extrinsic curvature tensor vanishes, and through
the Gauss-Codazzi relation also the intrinsic curvature scalar of the
time slice
vanishes. It is interesting that such $z_B$ coincide with the
positions of the accessory singularities which will appear in the
uniformization problem. These accessory parameters summarize all
the gravitational interaction.
An interesting restriction occurs on the conjugate momenta
i.e. $\sum_n P_n=0$ \cite{Des85}. In absence of such a restriction no solution
exists to eq.(\ref{liouvilleeq}) with $\exp({2\sigma})$ behaving at
infinity with 
$\exp({2\sigma}) \approx s^2 (z\bar z)^{-\mu}$ which describes a space
geometry asymptotic  to a cone. This can be easily seen
by applying the divergence theorem to the analogous of
eq.(\ref{liouvilleeq}) for 
$\sigma$. The same reasoning excludes the addition of an holomorphic
term to the $\pi^{\bar z}_{~z}$ given by eq.(\ref{pibarzz})

The solution of the inhomogeneous Liouville equation
(\ref{liouvilleeq}) can be understood as a variant of the Riemann-
Hilbert problem and is reduced to a linear fuchsian
differential 
equation \cite{bolibrukh}. By taking the ratio of two solutions of such
equation one can build the 
function $f(z)$ which maps the metric of the Poincar\'e pseudo-sphere
into the conformal factor which solves (\ref{liouvilleeq}) according to
the formula \cite{kra} 
\begin{equation}\label{fexpression}
e^{-2\tilde\sigma} =\frac{8 f'(z) \bar f'(\bar z)}{(1-f(z)\bar f(\bar
z))^2} = \frac{{\rm const}}{(\bar y_2(\bar z) y_2(z) - \bar y_1(\bar z)
y_1(z))^2};~f(z) =\frac{y_1(z)}{y_2(z)} 
\end{equation}
It is a variant of the Riemann-Hilbert problem as we are not given
directly with the monodromies but with the following information on
them: all monodromies belong to the $SU(1,1)$ group, otherwise the conformal
factor (\ref{fexpression}) is not single valued and as such cannot  
solve the Liouville equation eq.(\ref{liouvilleeq}); in addition we
are given with the conjugation classes of the monodromies around 
the particles (the particle masses), the conjugation class of the
monodromy at infinity (the total energy) and the positions of the
auxiliary singularities $z_B$, which are equivalent to the knowledge
of the particle momenta $P_n$ up to a multiplicative factor. Due to
the relation $\sum_nP_n=0$ the number of auxiliary singularities is
${\cal N}-2$. 

In order to count the
physical degrees of freedom in addition to the particle positions, we
must keep in mind that for ${\cal N}$ particles we have ${\cal N}$
$SU(1,1)$ monodromies to which we have to subtract a $SU(1,1)$
conjugation thus reaching $3{\cal N}-3$ real degrees of freedom. These
are equivalent to giving the ${\cal N}$ particle masses $\mu_i$, the
total energy $\mu$ and the complex positions $z_B$ of the ${\cal N}-2$
auxiliary singularities i.e. $3{\cal N}-3= {\cal N} +1 +2({\cal N}-2)$. The
linear residues $\beta_B$ of the  
auxiliary singularities are fixed by the solution of the Riemann-
Hilbert problem while the linear residues $\beta_i$ at the particle
singularities are computed from the ${\cal N}-2$ no logarithm
condition at the auxiliary singularities and the first and second
Fuchs conditions on the residues of the fuchsian differential
equation.

The conformal factor $\tilde\sigma$ is the key quantity in all the
subsequent developments. In fact a secondary constraint following from
the primary gauge constraint $K=0$ is
\begin{equation}\label{Nequation}
\Delta N = e^{-2\tilde\sigma} N
\end{equation}
and such $N$ can be computed from the knowledge of $\tilde
\sigma$. The reason is that the solution of the inhomogeneous Liouville
equation (\ref{liouvilleeq}) contains a free parameter $\mu$ which is
related to the  
behavior at infinity of the conformal factor i.e. $\exp(2\sigma) = s^2
(z\bar z)^{-\mu}$. As the sources do not depend on $\mu$ a solution of
eq.(\ref{Nequation}) is given by
\begin{equation}\label{N}
N = \frac{\partial(-2\tilde\sigma)}{\partial M}
\end{equation}
and one easily proves such a solution to be unique.
The other secondary constraint on $N^{z}$ is 
\begin{equation}
\partial_{\bar z}N^z=- \pi^z_{~\bar z}~e^{-2\sigma} N
\end{equation}
solved by
\begin{equation}\label{Nz}
N^z =-\frac{2}{\pi^{\bar z}_{\ z}(z)} \partial_z N +g(z).
\end{equation}
Here $g(z)$ is a meromorphic function whose role is to cancel the
poles occurring in
the first member on the r.h.s. due to the
zeros of $\pi^{\bar z}_{\ z}$.
The expression of $g(z)$ in $N^z$ is \cite{MS1}
\begin{equation}\label{generalg}
g(z) = \sum_B \frac{\partial\beta_B}{\partial M} \frac{1}{z-z_B} \frac{{\cal
P}(z_B)}{\prod _{C\neq B} (z_B-z_C)} + p_1(z). 
\end{equation}
$p_1(z)$ is a first order polynomial related to the motion of
the frame at infinity, more properly \cite{henneaux,ashtekar} to the
gauge transformation  
of translations, rotations and dilatations which leave invariant the
conformal structure of the space metric and leave fixed the point
at infinity.

In conclusion the metric is obtained in a straightforward way from
$\tilde\sigma$. 
The particle equations of motion
are extracted from the variation of the action with respect
to particle momenta and coordinates and take the form \cite{MS1}
$$
\dot z_n = - N^z(z_n) = -g(z_n) = 
$$
\begin{equation}\label{dotz}
-\sum_B
\frac{\partial\beta_B}{\partial M}\frac{1}{z_n-z_B} 
\frac{{\cal P}(z_B)}{\prod _{C\neq
B} (z_B-z_C)} - p_1(z_n)
\end{equation}
$$
\dot P_{n z} = 4\pi \frac{\partial \beta_n}{\partial M}+P_{nz} ~
g'(z_n)=
$$
\begin{equation}\label{dotP}
=4\pi \frac{\partial \beta_n}{\partial M} - P_{nz}\sum_B
\frac{\partial \beta_B}{\partial M} \frac{{\cal P}(z_B)}{(z_n-z_B)^2
\prod_{C\neq B}(z_B-z_C)}
+ P_{nz} ~p'_1(z_n).
\end{equation}
As already mentioned the linear polynomial $p_1(z)$ is related to the
choice of the frame at infinity. The choice of the frame which
does not rotate and dilatate at infinity  imposes that $N^z$ contains no
linear term and thus fixes 
\begin{equation} 
p_1(z) = c_0 -~\frac{1}{\sum_n P_n z_n} ~z.
\end{equation}
An interesting generalized conservation law which holds for any number
of particles can be derived \cite{MS1} from
eqs.(\ref{dotz},\ref{dotP}) 
\begin{equation}\label{dilconserv}
\sum_n P_n z_n = (1-\mu)(t-t_0) - iL. 
\end{equation}
\begin{figure}[b]
\includegraphics[width=.5\textwidth]{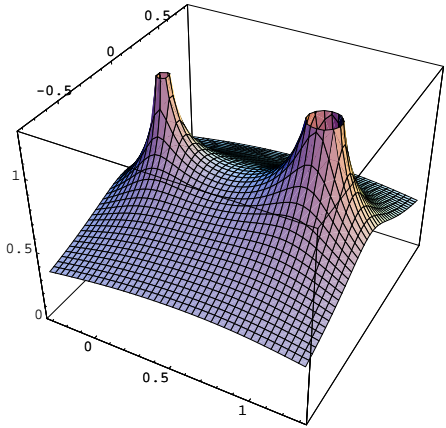}
\caption[]{Conformal factor for the classical two body problem}
\end{figure}

For the two body problem we have no auxiliary singularity  and the
fuchsian differential equation underlying the solution of
eq.(\ref{liouvilleeq}) is the hypergeometric equation. The conformal
factor can be easily expressed in term of 
hypergeometric functions, thus giving a complete information on the
metric.   
Equations (\ref{dotz},\ref{dotP}) take a very simple form in
the two body case due to the absence of apparent
singularities.  It is interesting that the equations of motion
can be  explicitly written just by looking at the local properties of
the fuchsian differential equation which underlies the solution of
eq.(\ref{liouvilleeq}). We obtain going over to the relative
coordinates $z'_2 
= z_2 - z_1$ and $P' = P_2 = - P_1$
\begin{equation}\label{system}
\dot z'_2 =\frac{1}{P'_z};~~~~\dot P'_z = - \frac\mu {z'_2}.
\end{equation}
being $4\pi \mu = M$ the total mass (energy) of the system.
One immediately realizes that such equations of motion (and their
complex conjugate) are generated by the hamiltonian 
$H = h + \bar h$ with $h= \ln(P' {z'}_2^\mu)$ where $h$ and $\bar
h$ are constants of motion. Taking the ratio of $P' {z'}_2^\mu = \exp(
h ) = {\rm const.}$ with eq.(\ref{dilconserv}) we obtain the solution
for the particle motion
\begin{equation}\label{solution}
z = {\rm const} [(1-\mu)(t-t_0) - iL]^{\frac{1}{1-\mu}}. 
\end{equation}
without the need to solve the system (\ref{system}). Thus the two body
problem is integrable. Solution
(\ref{solution}) was first found in \cite{BCV}. The fact that only the
total energy $M= 4\pi \mu$ intervenes in eq.(\ref{solution}) constitutes
a proof of a conjecture by 't Hooft \cite{hooft3} about the independence of the
solution of the two body problem of the masses of the constituent 
particles. 

In Fig.1 we report the conformal factor for the two body problem for
$\mu_1 = 0.1$, $\mu_2 = 0.2$ and $\mu = 0.5$


\section{The {\cal N} body problem}

In the coordinate frame in which the polynomial $p_1$ vanishes the
equations of motion in the relative coordinates $z'_n = z_n-z_1$,
$P'_n = P_n$ for $n=2\dots {\cal N}$ take the form \cite{CMS}
\begin{equation}\label{dotz1}
\dot z'_n = -\sum_B\frac{\partial\beta_B}{\partial\mu}\frac{\partial
z'_B}{\partial P'_n} ~~~~~~~~n=2,\dots {\cal N} 
\end{equation}
and
\begin{equation}\label{dotP1}
\dot P'_n = \frac{\partial\beta_n}{\partial\mu}+
\sum_B\frac{\partial\beta_B}{\partial\mu}\frac{\partial z'_B}{\partial
z'_n} ~~~~~~~~n=2,\dots {\cal N}. 
\end{equation}
These equations are canonically related to the ones with $p_1\neq 0$
and thus is sufficient to examine them.
The problem is now to show that such system is of hamiltonian nature;
such a result is expected as we obtained the above equations by
reduction of a hamiltonian system. Despite that, it is of interest to
have a direct proof of it and an expression of the hamiltonian.

In the simpler case of three body, where we have a single
auxiliary singularity, one immediately sees that the
hamiltonian has to be of the form
\begin{equation}\label{threehamiltonian}
H(z'_2,z'_3, P'_2,P'_3) = -\int_{z_0}^{z'_A}\frac{\partial
\beta_A}{\partial \mu}(z'_2,z'_3,z''_A) ~d z''_A + f(z'_2, z'_3) +~ c.c.
\end{equation} 
$H$ generates the equations for $\dot z'_n$ for any $f(z'_2, z'_3)$;
the function $f$ has to be determined by requiring that the same $H$
generates also the equations for $\dot P'_n$. This imposes some
integrability conditions on the function $\beta_A$; such conditions are
satisfied due to the validity of the Garnier equations which give a
constraint on the 
evolution of the auxiliary parameters under isomonodromic
deformations \cite{yoshida,okamoto}. The fact that our deformations
are isomonodromic is a 
consequence of the constancy in $2+1$ dimensional gravity of
the monodromies around the particle world lines. It is of interest
that such equations can be derived directly form the ADM formalism as
equations of motion \cite{MS1}.

The problem with four or more particle, when we are in presence of two
or more auxiliary singularities is more difficult and it related to an
interesting conjecture due to Polyakov \cite{ZT1} about the auxiliary
parameters 
of the $SU(1,1)$ Riemann-Hilbert problem. Such a conjecture states
that  the regularized Liouville action \cite{takh}   
$$
S_\epsilon [\phi] =\frac{i}{2} \int_{X_\epsilon} (\partial_z\phi 
\partial_{\bar z} \phi +\frac{e^\phi}{2}) dz\wedge d\bar z
-\frac{i}{2}\sum_n(1-\mu_n)\oint_n\phi(\frac{d\bar z}{\bar z -\bar
z_n}- \frac{d z}{ z - z_n})
$$
$$
+\frac{i}{2}\sum_B \oint_B\phi(\frac{d\bar z}{\bar z -\bar
z_B}- \frac{d z}{ z - z_B})
-\frac{i}{2}(\mu-2)\oint_\infty\phi(\frac{d\bar z}{\bar z}- \frac{d
z}{z})
$$
\begin{equation}\label{Sepsilon}
-\pi\sum_n(1-\mu_n)^2 \ln\epsilon^2 -\pi\sum_B\ln\epsilon^2 -\pi
(\mu-2)^2\ln\epsilon^2 
\end{equation}
is the generator of the linear residues $\beta_i$, $\beta_B$ in the
fuchsian differential  equation whose solutions provide the mapping
function which solve the inhomogeneous Liouville equation, i.e.
\begin{equation}\label{Sepsilonmu}
-\frac{1}{2\pi} d S_\epsilon = \sum_n\beta_n dz_n + \sum_B \beta_B dz_B+~c.c.  
\end{equation}
Such a conjecture has been proven by Zograf and Takhtajan \cite{ZT1} for
the case 
of parabolic singularities and elliptic singularities of finite
order. In the context of 2+1 dimensional gravity with generic masses
one should need the validity of such a conjecture for generic elliptic
singularities.  
The relevance of such a conjecture is that it gives a new meaning
to the auxiliary parameters; moreover it is a straightforward
consequence of eq.(\ref{Sepsilonmu}) that, apart from a constant
\begin{equation}\label{Smu}
H = \frac{1}{2\pi}\frac{\partial S_\epsilon}{\partial \mu}\
\end{equation}
as can be seen by computing $\frac{\partial H}{\partial P'_n}$,
$\frac{\partial H}{\partial z'_n}$ and comparing with
eqs.(\ref{dotz1},\ref{dotP1}). 
We recall that in the non rotating frame the hamiltonian
contains an additional contribution, as already observed in
sect. 3. Its complete form in that  
frame is indeed given by \cite{CMS}
\begin{equation}\label{eqhamiltonian}
H= \ln\left[(\sum_n P_n z_n) (\sum_n \bar P_n \bar z_n)\right] +
\frac{1}{2\pi}\frac{\partial S_\epsilon}{\partial \mu}. 
\end{equation} 
Note that this hamiltonian, being time-independent,  provides a further
conservation law in the ${\cal N}-$particle problem.

The hamiltonian $H$ has been constructed from the equations of
motion. On the other hand it should be possible to derive directly the
reduced hamiltonian starting from the action after replacing into it
the solution of the constraints. As in our gauge $\pi^{ij}\dot
g_{ij}=0$, the action reduces to
\begin{equation}
S  = \int dt (\sum_n P_{ni}\, \dot q_n^i -H_B)
\end{equation}
with $H_B$ given by eq.(\ref{boundaryhamiltonian}). The last term in
$H_B$ contributes to 
a constant while the term proportional to $\eta/\cosh\eta$ goes to
zero at large distances. Thus we are left only with the term
proportional to $K_{Bt}$ which for large $r_0$ can be computed to give
\cite{CMS} 
\begin{equation}\label{HBr0}
H_B = - r_0\ln r_0^2 (\frac{1}{r_0}+ \partial_r \sigma)=
(\mu-1) \ln r_0^2. 
\end{equation} 

We recall now that the equations of motion are obtained from the
action by keeping the values of the fields fixed at the boundary, or
equivalently \cite{wald} by keeping fixed the intrinsic metric of
the boundary. In our case the variations should be performed keeping
fixed the fields $N$, $N^a$, and $\sigma$ at the boundary.
We shall perform the
computation for the boundary given by a circle of radius $r_0$ for a
very large value of $r_0$.  
The asymptotic form of the conformal factor $2\sigma$ is 
\begin{equation}
2\sigma = \ln s^2 -\mu \ln r_0^2.
\end{equation} 
If we change the particle positions and momenta, $s^2$
varies and in order to keep the value of $\sigma$ fixed at the
boundary we must vary $\mu$ as follows
\begin{equation}
0= - \delta \mu \ln r_0^2 + \delta\ln s^2 
\end{equation} 
i.e. for large $r_0$
\begin{equation}
\delta \mu \approx \frac{1}{\ln r_0^2}\delta \ln s^2.
\end{equation} 
Substituting into eq.(\ref{HBr0}) 
we have 
\begin{equation}\label{shamiltonian}
H_B = \ln s^2 +{\rm const.} 
\end{equation}
i.e. the hamiltonian is given by the logarithm of the coefficient of
the asymptotic expansion of the conformal factor at large distances.

We want now to relate the result eq.(\ref{shamiltonian}) to the
results obtained directly from the equations of motion.

Let us consider the value of the
action $S_\epsilon$ on the solution of the Liouville
equation and let 
us compute its derivative with respect to $\mu$. As we are varying
around a stationary point the only contribution is provided by the
terms in eq.(\ref{Sepsilon}) which depend explicitly on $\mu$ i.e.
\begin{equation}
\frac{\partial S_\epsilon}{\partial \mu} = -\frac{i}{2}
\oint_\infty\phi(\frac{d\bar z}{\bar z}- \frac{d z}{ z}) - 2\pi
(\mu-2)\ln\epsilon^2  
\end{equation}
and as $\phi\equiv -2\tilde\sigma$ at infinity behaves like
\begin{equation}
\phi\approx (\mu-2) \ln z\bar z +\ln\left[(\sum_n P_n z_n) (\sum_n \bar
P_n \bar z_n)\right] -\ln s^2 +{\rm const} 
\end{equation}
we have
\begin{equation}
\frac{1}{2\pi}\frac{\partial S_\epsilon}{\partial \mu} =
-\ln\left[(\sum_n P_n z_n) (\sum_n \bar P_n \bar z_n)\right] +\ln s^2
+{\rm const} 
\end{equation}
which substituted in eq.(\ref{shamiltonian}) gives 
\begin{equation}
H = \ln\left[(\sum_n P_n z_n) (\sum_n \bar P_n \bar z_n)\right] +
\frac{1}{2\pi}\frac{\partial S_\epsilon}{\partial \mu} + {\rm const}
\end{equation}
in agreement with the result (\ref{eqhamiltonian}) obtained through
Polyakov's conjecture.

It is remarkable that the same hamiltonian is obtained independently
from the boundary term of the $2+1$ gravitational action and from the
equations of motion in conjunction with Polyakov conjecture. This
lends support to the validity of Polyakov's conjecture in the general
elliptic case or at least to a weak form of it, obtained by taking the
derivative of eq.(\ref{Sepsilonmu}) with respect to $\mu$.

\section{Quantization: the two particle case}

\begin{figure}[b]
\includegraphics[width=.5\textwidth]{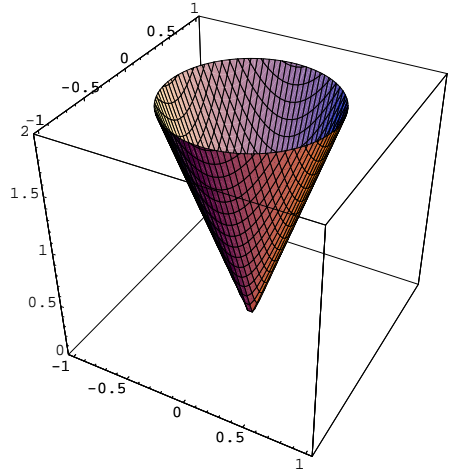}
\caption[]{Cone metric equivalent to the classical metric of Fig.1,
intervening in the quantum two body problem}
\end{figure}

Several quantization scheme have been put forward in the context of
$2+1$ dimensional gravity (see
e.g. \cite{hosoya2,carlip,hooft,wael,nelsonregge,matschull}). Here we shall examine 
the canonical quantization which flows directly from the hamiltonian
ADM formalism \cite{ADM}. 
We recall that the classical two particle hamiltonian in the reference 
system which does not rotate at infinity is given by
\begin{equation}
H = \ln(P z \bar P \bar z) + (\mu -1)\ln (z\bar z)=
\ln(Pz^\mu) + \ln(\bar P\bar z^\mu) = h + \bar h
\end{equation}
with $P = P'_2$ and $z = z'_2$.
$H$ can be rewritten in cartesian coordinates as
\begin{equation}
H = \ln((x^2+y^2)^\mu ((P_x)^2 + (P_y)^2)).
\end{equation}
Keeping in mind that with our definitions $P$ is the momentum
multiplied by $16\pi G_N/c^3$, applying the correspondence principle 
we have
\begin{equation}
[x,P_x] = [y,P_y] = i l_{P}
\end{equation}
where $l_P = 16 \pi G_n\hbar/c^3$,
all other commutators equal to zero. $H$ is converted into the
operator
\begin{equation}\label{logbeltrami}
\ln[-(x^2+y^2)^\mu \Delta] +~{\rm constant}.
\end{equation}
The argument of the logarithm is the Laplace-Beltrami
$\Delta_{LB}$ operator on the metric $ds^2=(x^2+y^2)^{-\mu} (dx^2 +
dy^2)$. We note that in the quantum problem only the simple metric of
a cone (see Fig.2) intervenes and not the classical metric of
Fig.1. The angular 
deficit of the cone again in given by the total energy thus proving 't
Hooft conjecture \cite{hooft3} at the quantum level. Following 
an argument similar to the one presented in   
\cite{sorkin} one easily proves that if we start from the 
domain of $\Delta_{LB}$ given by the infinite differentiable functions
of compact support $C^\infty_0$ which can also include the origin,
then $\Delta_{LB}$ has a unique 
self-adjoint extension in the Hilbert space of functions square
integrable on the metric $ds^2=(x^2+y^2)^{-\mu} (dx^2+dy^2)$
\cite{CMS} and as a
result since $\Delta_{LB}$ is a positive operator, $\ln(\Delta_{LB})$ is
also self-adjoint. 

Deser and Jackiw \cite{deserjackiw} considered the quantum scattering
of a test particle 
on a cone both at the relativistic and non relativistic level. Most of
the techniques developed there can be transferred here. The main
difference is the following; instead of the 
hamiltonian $(x^2+y^2)^\mu(p_x^2+p_x^2)$ which appears in their non
relativistic treatment, we have now the hamiltonian
$\ln[(x^2+y^2)^\mu(p_x^2+p_y^2)]$. The partial wave eigenvalue
differential equation 
\begin{equation}
(r^2)^\mu[-\frac{1}{r} \frac{\partial }{\partial r} r \frac{\partial
}{\partial r}+\frac{n^2}{r^2} ]\phi_n(r) = k^2 \phi_n(r) 
\end{equation}
with $\mu=1-\alpha$ is solved by
\begin{equation}
\phi_n(r) = J_\frac{|n|}{\alpha}(\frac{k}{\alpha}r^\alpha)
\end{equation}
and out of them one can compute the quantum Green function which in
our case can be expressed in terms of hypergeometric functions
\cite{CMS}. The ordering problem always subsists; the one adopted in
(\ref{logbeltrami}) is the most natural. In the ${\cal N}$-body
problem due to the complexity of the hamiltonian
we expect the ordering problem to be more acute. Power expansions in
some small parameter like the total kinetic energy or the particle
masses may give useful indications. 

\bigskip
\bigskip
\bigskip


\section*{Acknowledgments}
This talk is based on papers written by the author in
collaboration with Domenico Seminara and Luigi Cantini to whom the
author is very grateful. The author is also grateful to Marcello
Ciafaloni, Stanley Deser and Roman Jackiw for useful discussions.

%

\end{document}